\documentclass[prl,twocolumn,amsmath,showpacs]{revtex4}
\input {epsf.sty}
\usepackage{epsfig}
\usepackage{amssymb}
\begin{document}
\title{Specific Heat of Disordered Superfluid $^{3}$He}
\author{H. Choi, K. Yawata, T.M. Haard, J.P. Davis, G. Gervais, N. Mulders, P. Sharma, J.A. Sauls, and W. P.
Halperin}
\affiliation{Department of Physics and Astronomy,\\
   Northwestern University, Evanston, Illinois 60208}

\date{Version \today}

\begin{abstract} The specific heat of superfluid $^{3}$He, disordered by a silica aerogel, is found to have a sharp
discontinuity marking the thermodynamic transition to superfluidity at a temperature reduced from that of bulk
$^{3}$He.  The magnitude of the discontinuity is also suppressed. This disorder effect can be understood from the
Ginzburg-Landau theory which takes into account elastic quasiparticle scattering suppressing both the transition
temperature and the amplitude of the order parameter.  We infer that the limiting temperature dependence of the specific
heat is linear at low temperatures in the disordered superfluid state, consistent with predictions of gapless excitations
everywhere on the Fermi surface.
\end{abstract}

\maketitle

\vspace{11pt}

Two essential characteristics of superconductors are manifest in the specific heat. First,
there is a jump at the transition temperature of magnitude that depends on the strength of the electron pair
interactions that lead to superconductivity.  Secondly, the temperature dependence at low temperature gives a
fingerprint of the energy gap structure.  Consequently, the measurement of the specific heat is one of the
most fundamental to understanding the nature of superconductivity.  In fully gapped superconductors, such as
aluminum, the specific heat decreases exponentially with decreasing temperature following an Arrhenius
relation\cite{Phi59}, thereby demonstrating unambiguously the existence of a full gap in the electron quasiparticle
excitation spectrum and providing a direct measurement of its size.  Both of these basic thermodynamic behaviors
conform to predictions of the Bardeen, Cooper, and  Schrieffer theory\cite{Tin96} and can be expected to hold in
general for any condensate of fermion pairs into a state that is fully gapped.  This is precisely what is
expected\cite{Leg75} and observed\cite{Gre86} for the B-phase of superfluid
$^{3}$He, a  {\it p}-wave superfluid with an isotropic energy gap,  similar to conventional
{\it s}-wave superconductors.  

Shortly after the success of BCS theory it was sought to understand how perturbations,
such as impurities, modify superconducting behavior.  It was found\cite{Abr59} that magnetic scattering of
quasiparticles suppresses both the transition temperature and the gap magnitude leading to `gapless
superconductivity'\,\cite{Tin96}.  The same should be true for all superconductors and fermion
superfluids even those that do not have {\it s}-wave states in which case all forms of  elastic scattering
will produce these suppression effects\cite{Abr61}.  In particular it should be true for superfluid $^{3}$He.  In
this letter we present systematic measurements of the heat capacity in the  B-phase of superfluid $^{3}$He,
disordered by  impurities, demonstrating suppression of the transition temperature, suppression of the order
parameter, and gapless superfluid behavior.

Superfluid $^{3}$He is a unique example of a Cooper pair condensation.  It was the first unconventional pairing
state discovered. By unconventional we mean that there are symmetries of the normal Fermi liquid
that are spontaneously broken in the superfluid phases in addition to gauge symmetry, e.g. spin and spatial rotations.  In this
sense $^{3}$He is similar to cuprate superconductors and to the heavy fermion superconductor UPt$_{3}$ as well as a number of
other recently discovered superconducting compounds.  However, in contrast to the superconductors, superfluid
$^{3}$He is a neutral system, easily prepared as the purest known material, whose general
properties  are very well established.  It is not so straightforward, however,  to
introduce impurity scattering in a controlled way in $^{3}$He.

Porto {\it et al.}\cite{Por95} and Sprague {\it et
al.}\cite{Spr95,Spr96}  found that they could use very dilute silica aerogels to introduce impurity scattering in
superfluid
$^{3}$He and subsequently there has been a wealth of new information about the nature of this modified superfluid
state.  Impurity scattering models for $^{3}$He were developed by Thuneberg {\it et al.}\cite{Thu98} with clear
predictions for the suppression of the transition temperature and the heat capacity
discontinuity.  Sharma and Sauls\cite{Sha01} have calculated transport behavior as well as the density of
quasiparticle excitations inside the gap which define its gapless superfluid behavior.  Our experimental results are
quantitatively consistent with the calculations of these basic thermodynamic properties.

Earlier work of He {\it et al.}\cite{He02} and preliminary results from our laboratory\cite{Yaw03} have provided
convincing evidence for the existence of a heat capacity anomaly at the transition temperature.  Additionally,
Fisher {\it et al.}\cite{Fis03} found that the thermal conductivity in the disordered A-superfluid state in a
magnetic field has a linear temperature dependence when extrapolated to  very low tempertature.  
 If the heat current were carried by gapless excitations at low energy there should be evidence for these excitations in
the low temperature specific heat.

\begin{figure}[b]
\centerline{\epsfxsize1.05\hsize\epsffile{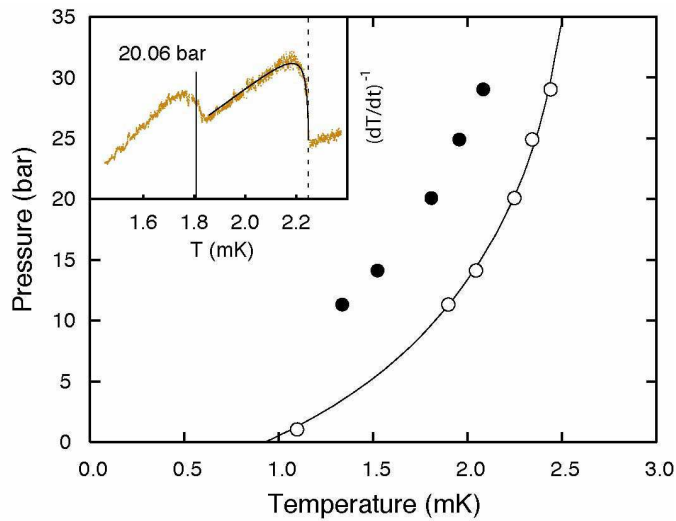}}
\begin{minipage}{0.93\hsize}
\caption {Pressure temperature phase diagram for bulk
$^{3}$He and disordered $^{3}$He in aerogel.  The smooth  curve is an interpolation
from Greywall's data\cite{Gre86} for bulk  $^{3}$He.  The data points are high resolution measurements of the
transition temperatures obtained during slow warming shown in the inset at $P=20.06$ bar.  Two heat capacity
anomalies are evident.   The solid fit curve  defines the bulk transition (dotted vertical
 line, open circles).  The  superfluid transition temperature in aerogel is at a lower temperature (solid vertical
line, solid circles), 
$\sim 80 \mu$K wide.}
\end{minipage}
\end{figure}
\noindent

Our low temperature calorimeter was cooled with a PrNi$_{5}$ adiabatic nuclear demagnetization refrigerator to
temperatures below 0.75 mK above which we made measurements of heat capacity using an adiabatic calorimetric
method.  Temperature was determined from measurement of the ac susceptibility of a lanthanum diluted
 cerium magnesium nitrate salt, calibrated with respect to a $^{3}$He melting curve thermometer according to the
Greywall temperature scale\cite{Gre86}.  The calorimeter volume was measured to be $1.91 \pm 0.02$ cm$^{3}$ with the
aerogel sample in place, leaving 44 \% of the space available for bulk $^{3}$He. The  porous silica aerogel was in the form
of a right circular disk
$0.380$ cm thick and
$1.902$ cm in diameter.  From its dry weight, 42.4 mg, we found the
porosity to be 98.2\%.  The sample was first cooled by nuclear demagnetization.  By opening a cadmium
superconducting heat switch we could isolate the sample cell from the nuclear stage, allowing the $^{3}$He to warm up
slowly owing to the ambient heat leak, typically $\dot{Q} \sim 0.1$ nW, starting in the B-phase\cite{Ger02}. Heat
pulses were applied with
$\Delta Q$ in the range $ \,0.1 \sim 0.5\, \mu $J  for $ 30 \sim 60$ s, and the temperature increments
$\Delta T$ were measured to obtain the heat capacity $C_{tot} = \Delta Q / \Delta T$ shown in Fig. 2 after
subtraction of the addendum.  A warm-up trace without heat pulses, in the form
$\dot{T}^{-1} = C/ \dot{Q}$, is shown in the inset to Fig.1 providing our highest resolution of both
transition temperatures, in bulk liquid $^{3}$He and $^{3}$He in aerogel.

We assume that liquid $^{3}$He above its transition to superfluidity
is a Fermi liquid with known interaction parameters even within the aerogel.  This is reasonable since the aerogel structure is
on a much larger scale than the Fermi wavelength\cite{Sau03}. Using the specific heat measurements by Greywall\cite{Gre86} on
bulk
$^{3}$He we can deduce from our experiments, a) the calorimeter background at temperatures above the 
superfluid transition in aerogel, b) the volumes of $^{3}$He in the aerogel and in the bulk, c) the heat
capacity jump in the disordered superfluid phase, and d) the temperature dependence of the heat capacity in the
disordered superfluid state.

The background heat capacity, $C_{add}$, can be entirely attributed to solid
$^{3}$He on the surface of the aerogel.  Two layers of $^{3}$He are 
paramagnetic\cite{Spr96,Bar00} and can be removed by substitution with two layers of
non-magnetic
$^{4}$He reducing the addendum to less than
$0.25\,
$mJ/K.  The helium isotope mixture experiments were limited to
$\geq$ 4 mK owing to long thermal time constants in the calorimeter.  The temperature and pressure dependence of the
addendum is shown in Fig. 3, of a comparable  magnitude\cite{Cbd} to that reported in vycor glass\cite{Gol96}.

\begin{figure}[!]
\centerline{\epsfxsize1.05\hsize\epsffile{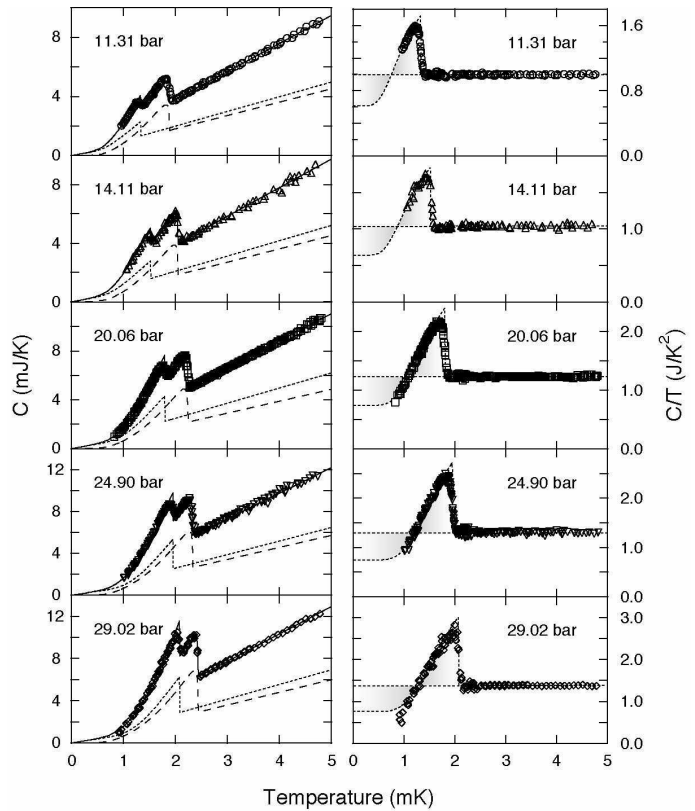}}
\begin{minipage}{0.93\hsize}
\caption {Heat capacity of  $^{3}$He in aerogel. On the left side we show the heat capacity as a function of
temperature for bulk $^{3}$He and
$^{3}$He in a 98\% porous silica aerogel contained together in the same calorimeter with addendum subtracted.
 On the right side we show the heat capacity divided by the
temperature for the disordered superfluid alone.  The curves represent the known temperature dependence of the
heat capacity for bulk
$^{3}$He (dashed) and a smooth fit to the data in the disordered superfluid with the constraint for entropy
conservation (dotted) described in the text.  }
\end{minipage}
\end{figure}
\noindent

The measured heat capacity, $C_{tot}$, corrected for the addendum, $C_{add}$, is 
\begin{equation}
{C(T,P)=C_{tot} - C_{add} = V_{a}c_{a} + V_{B}c_{B}}
\end{equation}
\noindent
shown in Fig. 2. 
Above the bulk transition temperature, $T_{c}$, $C(P,T)$ is proportional to  temperature, since the specific heats of
the fluid inside, $c_{a}$, and outside, $c_{B}$, the aerogel are identical:  $c_{a} = c_{B}=\gamma T$ where $\gamma =
(\pi^{2}/3)k_{B}^{2}N_{0}$ and $N_{0}$ is the density of states at the Fermi energy. Knowing the total volume
$V_{a}+V_{B}$ of aerogel and bulk fluids, we can determine the addendum in this temperature range.  At $T_{c}$ the
bulk heat capacity jump was compared with Greywall's values to give a direct measure of $V_{a}$ from which we found
the fluid volume in the aerogel to be $1.028 \pm 0.021$cm$^{3}$, consistently at all pressures and  close\cite{Bun00} to
the measured geometry of the aerogel giving a pore volume of
$1.062
\pm 0.006$ cm$^{3}$.  With $V_{a}$ and Eq. 1, we can determine the addendum in the range
$T_{c} > T > T_{c,a}$, by comparing our measurements with the known specific heat of the bulk superfluid\cite{Gre86}.  We must
also allow for contributions to the heat capacity from a normal state region at the surface of the heat exchanger,
approximately  a coherence length,
$\xi(T)$, in depth.  This small contribution is found by fitting the slow warm-up curves as shown in the inset to Fig.1 to
obtain an effective area of the heat exchanger $A = 1.71$m$^{2}$, valid at all pressures.  The temperature dependence of the
addendum at 1.02 bar is smooth to below 1 mK.  Consequently, it is reasonable to assume that this is the case at
higher pressures and to extend linearly the temperature dependence of the addendum from $T_{c,a}$ to lower temperatures in
order to extract the heat capacity,
$C_{a}$, of the disordered superfluid $^{3}$He, Fig.2.  

 We show $C_{a}/T$ on the right side of Fig.2 
after subtracting the bulk contribution, the surface contribution, and the addendum.  The heat capacity jumps
appear quite sharp in the figure but substantially reduced even compared to the weak coupling BCS value of
$\Delta C/C =1.43$.   They are shown as a function of pressure in Fig. 4, normalized to their bulk values. 
Consequently, the amplitude of the order parameter is also reduced since the jump is related to the
temperature dependence of the gap function,
$\Delta_{a}(T)$, near
$T_{c,a}$, in the Ginzburg-Landau limit,   

\begin{equation}
{\Delta_{a}(T) = \pi k_{B} T_{c,a}\Bigl({2 \over 3} {\Delta C_{a} \over C_{a}}{{T_{c,a} - T \over
T}}\Bigr)^{1/2}} \quad T \lesssim T_{c}
\end{equation}
\noindent

\begin{figure}[!]
\centerline{\epsfxsize0.93\hsize\epsffile{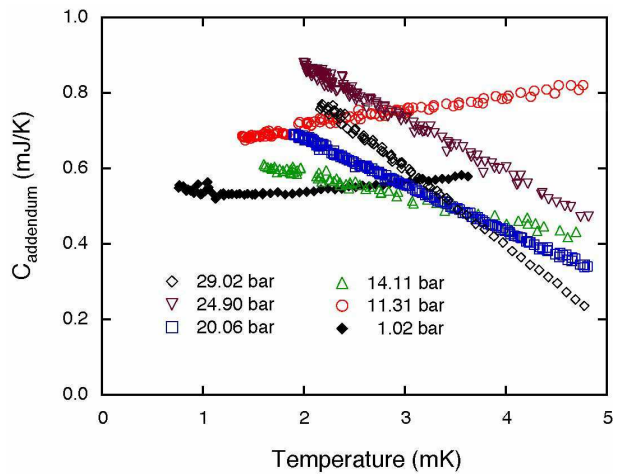}}
\begin{minipage}{0.93\hsize}
\caption {Heat capacity addendum.  The temperature and pressure dependence of the heat capacity addendum is shown for
bulk 
$^{3}$He in aerogel with an estimated accuracy of 0.25 mJ/K. }
\end{minipage}
\end{figure}
\noindent

At lower temperatures, below $T_{c,a}$, we note that 
$C_{a}/T$ appears to be quite linear.  We take advantage of this fact and we invoke the third law of thermodynamics, 

\begin{equation}
{0 = \int_0^{T_{c,a}} (\gamma - C_{a}/T) \,dT},
\end{equation}
\noindent
to determine the limiting behavior of $C_{a}$ as $T \rightarrow 0$ shown by dotted curves in Fig. 2.  Graphically,
this means that for each pressure the two shaded areas must be equal.  If the heat
capacity is a monotonic function of temperature, as is expected theoretically, then to satisfy Eq. (3), 
$C_{a}/T$ must have a non-zero intercept at $T=0$ requiring a non-zero density of states, $N_{a}(0)$, at the Fermi
energy in the superfluid state. The curves describing this behavior in the disordered superfluid were constructed
phenomenologically to satisfy the constraints and our interpretation does not depend on its functional form nor is
it particularly sensitive to the data at the lowest temperatures.  For example in Fig.2  at $P= 29.02$ bar the data
fall below the curve likely owing to our overestimate of the addendum. The low energy
excitations affecting the low temperature specific heat are a broad band of Andreev  bound states, centered at the Fermi
energy, that form near the aerogel strands\cite{Sha01}. As a result superfluid $^{3}$He-B in aerogel is a `gapless
superfluid'.   

\begin{figure}[!]
\centerline{\epsfxsize1.0\hsize\epsffile{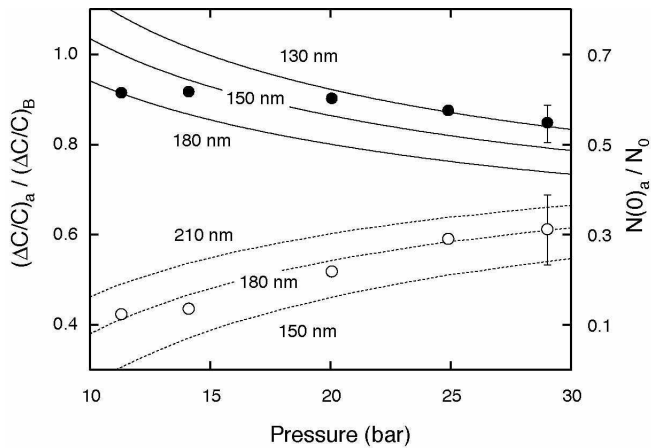}}
\begin{minipage}{0.93\hsize}
\caption {Heat capacity jump and density of states.  The pressure dependence of the heat capacity
jump (open circles, left axis) and the density of states (filled circles. right axis) at zero energy (the Fermi
energy) is shown for bulk 
$^{3}$He in a 98\% porous silica aerogel.  The error bars arise from possible systematic error in the
choice of temperature dependence for the heat capacity just below $T_{c,a}$.  Other statistical errors are of the size of
the data points.  Calculations from the HISM are shown as curves for various values of the transport mean free path.}
\end{minipage}
\end{figure}
\noindent

The simplest theoretical description\cite{Thu98, Sha01} considers isotropic quasiparticle scattering spread
homogeneously throughout the aerogel volume, characterized by a single parameter, the transport-mean-free path,
$\lambda_{T}$. This model (HISM) provides a qualitatively
consistent  account of the superfluid transition temperature suppression\cite{Thu98}, the suppression of the A-phase
and the polycritical point\cite{Ger02,Sau03},  measurements of spin diffusion\cite{Col02} and the thermal
conductivity\cite{Sha01,Fis03}.  Here we see that the model gives a quantitatively consistent picture 
for the  heat capacity jump over a wide
range of pressure. A calculation\cite{Thu98} with the HISM gives a best fit value of
$\lambda_{T} = 180$ nm. In this fit we rescale strong
coupling contributions\cite{Ger02} to the free energy by the ratio $T_{c,a}/T_{c}$ where $T_{c,a}$ is calculated
self-consistently within the HISM.  We also find reasonable agreement with the prediction from the HISM for the
density of states of gapless excitations.  The calculations as a function of pressure for $N_{a}(0)$ are shown in
Fig.4 for several values of
$\lambda_{T}$ near
$150$ nm and are qualitatively consistent with our analysis of the heat capacity jump, $\lambda_{T} = 180$ nm.

Discrepancy between the experiment and the model for the density of states can be attributed in part to estimated
systematic errors including the   extrapolation of the addendum to $T \leq T_{c,a}$ and possible anisotropic or
inhomogeneous scattering.  Nonetheless, the  consistency that is apparent within the heat capacity experiment 
provides strong evidence that elastic scattering suppresses superfluidity and is approximately  isotropic and
homogeneous.  It is interesting to compare our results with unconventional superconductors in several cases. 
Precision measurements of the transition temperature of UPt$_{3}$ indicate\cite{Kyc98}  suppression of the transition
temperature with increased scattering but, in this case, with a significant  anisotropy.  In cuprates doped with
Zn impurity, such as  YBa$_{2}$(Cu$_{1-y}$Zn$_{y}$)$_{3}$O$_{7}$, suppression effects have been observed in
the heat capacity \cite{Lor94}, strikingly similar to those we report here.

We acknowledge support from the National Science Foundation DMR-0244099 and we thank Ryuji Nomura, and
John Halpine for assistance in the early stages of this project and helpful conversations with Yoonseok Lee, John
Reppy, and Jeevak Parpia.


\begin{thebibliography}{70}
\bibitem{Cbd} The aerogel surface area is 45 m$^{2}$ from helium BET
measurements on a similar sample giving a surface solid specific heat of $11 \,\mu $J/Km$^{2}$ near 4 mK similar to
that found by Golov and Pobell\cite{Gol96} for $^{3}$He on vycor glass, $13 \,\mu $J/Km$^{2}$.
\end{thebibliography}
\end{document}